\RequirePackage{rotating}
\documentclass[acmsmall,screen]{acmart}

\usepackage[utf8]{inputenc} % allow utf-8 input
\usepackage[T1]{fontenc}    % use 8-bit T1 fonts
\usepackage{hyperref}       % hyperlinks
\usepackage{url}            % simple URL typesetting
\usepackage{amsfonts}       % blackboard math symbols
\usepackage{nicefrac}       % compact symbols for 1/2, etc.
\usepackage{microtype}      % microtypography

\usepackage{color}
\usepackage{multirow}
\usepackage{bm}
\usepackage{rotating}
\usepackage[lined, ruled, commentsnumbered, linesnumbered]{algorithm2e}
\usepackage{subcaption}
\SetKwRepeat{Do}{do}{while}%

\usepackage{amssymb}
\usepackage{enumitem}
\setlist[itemize]{leftmargin=*}

\definecolor{gray}{RGB}{192,192,192}
\definecolor{MyPink}{RGB}{255,178,178}
\definecolor{MyBlue}{RGB}{178,178,255}

\copyrightyear{2024}
\acmYear{2024} 
\setcopyright{acmlicensed}
\acmPrice{}
\begin{document}

\title{PAP-REC: Personalized Automatic Prompt for Recommendation Language Model}

\author{Zelong Li}

\affiliation{
  \institution{Rutgers University}
  \city{New Brunswick, NJ, US}
  \country{USA}
}
\email{zelong.li@rutgers.edu}

\author{Jianchao Ji}
\affiliation{
  \institution{Rutgers University}
  \city{New Brunswick, NJ, US}
  \country{USA}
}
\email{jianchao.ji@rutgers.edu}

\author{Yingqiang Ge}
\affiliation{
  \institution{Rutgers University}
  \city{New Brunswick, NJ, US}
  \country{USA}
}
\email{yingqiang.ge@rutgers.edu}

\author{Wenyue Hua}
\affiliation{
  \institution{Rutgers University}
  \city{New Brunswick, NJ, US}
  \country{USA}
}
\email{wenyue.hua@rutgers.edu}

\author{Yongfeng Zhang}
\affiliation{
  \institution{Rutgers University}
  \city{New Brunswick, NJ, US}
  \country{USA}
}
\email{yongfeng.zhang@rutgers.edu}

\def\authors{Zelong Li, Jianchao Ji, Yingqiang Ge, Wenyue Hua, Yongfeng Zhang}

\begin{abstract}

Recently emerged prompt-based Recommendation Language Models (RLM) can solve multiple recommendation tasks uniformly. The RLMs make full use of the inherited knowledge learned from the abundant pre-training data to solve the downstream recommendation tasks by prompts, without introducing additional parameters or network training. 
However, handcrafted prompts require significant expertise and human effort since slightly rewriting prompts may cause massive performance changes. In this paper, we propose \textbf{PAP-REC}, a framework to generate the Personalized Automatic Prompt for RECommendation language models to mitigate the inefficiency and ineffectiveness problems derived from manually designed prompts. Specifically, personalized automatic prompts allow different users to have different prompt tokens for the same task, automatically generated using a gradient-based method. One challenge for personalized automatic prompt generation for recommendation language models is the extremely large search space, leading to a long convergence time. To effectively and efficiently address the problem, we develop surrogate metrics and leverage an alternative updating schedule for prompting recommendation language models.
Experimental results show that our PAP-REC framework manages to generate personalized prompts, and the automatically generated prompts outperform manually constructed prompts and also outperform various baseline recommendation models. The source code of the work is available at \url{https://github.com/rutgerswiselab/PAP-REC}.

\end{abstract}

% <concept_id>10002951.10003317.10003338.10003341</concept_id>
% <concept_desc>Information systems~Language models</concept_desc>
% <concept_id>10002951.10003317.10003347.10003350</concept_id>
% <concept_desc>Information systems~Recommender systems</concept_desc>

% \ccsdesc[500]{Information systems~Language models}
% \ccsdesc[500]{Information systems~Recommender systems}

% \balance


\maketitle

\section{Introduction}
\label{sec:introduction}

As humans gradually enter the age of information overload, finding effective and personalized resources among the vast amount of information has become a crucial challenge, where recommendation systems (RS) have emerged as a solution. According to users' different needs and usage scenarios, various recommendation tasks also arise, including sequential recommendation, explainable recommendation, user-item matching, etc.

For a long period of time, RS models, from factorization machines \cite{rendle2010factorization} and gradient-boosted decision trees (GBDT) \cite{friedman2001greedy, he2014practical, ke2017lightgbm} to various deep neural networks \cite{lecun2015deep, cheng2016wide, zhang2019deep}, are designed and trained separately for each task in order to achieve high performance. However, recently proposed recommendation language models (RLM), such as P5 \cite{geng2022recommendation} and M6-Rec \cite{cui2022m6}, have revolutionized the existing training paradigm for RS. The RLMs are fined-tuned from language models that contain strong inherited knowledge learned from pre-trained data and convert various recommendation tasks into language understanding and generation problems. Thus, the RLM can solve multiple downstream tasks with different prompts without additional parameter learning or neural network training, which is a more effective and efficient way for task transfer.

Although the RLM has achieved performance improvement in various recommendation tasks, it relies on manually designed prompts, which require significant human efforts and may not achieve optimal results \cite{liu2021pre}. Recent research has shown that rewriting the prompt can lead to huge performance changes \cite{reynolds2021prompt}. Thus, reducing human efforts and pursuing optimal prompts are two motivations for designing an automatic prompt-generation framework.

In the field of natural language processing (NLP), there have been several research efforts that are related to automated prompt learning, including continuous prompt learning \cite{zhang2021differentiable, hambardzumyan2021warp, li2021prefix, liu2021gpt, zhong2021factual, zhou2022learning} and discrete prompt learning \cite{jiang2020can, autoprompt:emnlp20, yuan2021bartscore, gao2021making, 10.1162/tacl_a_00468, zhang2022promptgen, deng2022rlprompt, Zhu2022PointCLIPV2, zhang2023automatic}. Our PAP-REC framework is inspired by gradient-based discrete prompt learning techniques \cite{wallace-etal-2019-universal, autoprompt:emnlp20}, but different from existing prompt learning research on two perspectives. First of all, our PAP-REC can generate personalized prompts, i.e., different users could have different prompt tokens in the same task. It is essential since personalization is one of the key characteristics of RS. However, personalized prompts inflate the search space since the number of users is much more than the tasks. To handle the challenge, we develop an iterative and alternative token update schedule. 
Secondly, due to the use of specialized metrics such as Hit Rate (HR) @ $k$ \cite{burke2005segment} or Normalized Discounted Cumulative Gain (NDCG) @ $k$ \cite{jarvelin2002cumulated, buttcher2016information}, RS tends to focus on the top-$k$ results rather than the top-$1$. Thus, existing gradient-based prompt learning methods, such as optimization along loss function gradient, may lead to the ineffectiveness of RLMs. To efficiently address the problem, we propose surrogate evaluation metrics for the token selection criterion on prompt generation. 
This paper makes the following key contributions:
\begin{itemize}
    \item We propose the PAP-REC framework, which can automatically generate personalized prompts for the RLM to enhance the recommendation performance.
    \item We develop personalized prompt learning and design an iterative and alternative token update schedule to solve the inflating search space derived from user-specific tokens.
    \item Specific to the metrics of RS, we develop surrogate metrics for PAP-REC to generate effective automated prompts efficiently.
    \item We conduct experiments on three real-world datasets to show that with the prompts generated from the PAP-REC framework, the RLM has great performance improvement in most cases.
\end{itemize}

In the following part of this paper, we first review the related work in Section \ref{sec:related_work}. In Section \ref{sec:problem}, we formalize the prompt generation problem and different recommendation tasks used in this paper. In Section \ref{sec:generation_process}, we introduce the detailed design of the prompt generation process along with methods for better generation effectiveness and efficiency. We provide and analyze the experimental results in Section \ref{sec:experiment}, and finally conclude our work and suggest potential avenues for future research in Section \ref{sec:conclusions}.

\section{Related Work}
\label{sec:related_work}
 
\subsection{Recommendation Language Model}
\label{sec:foundation_model}

Pre-trained language models (LM) such as GPT series \cite{radford2019language, brown2020language, openai2023gpt4}, T5 series \cite{raffel2020exploring, sanh2021multitask, chung2022scaling}, and LLaMA series \cite{touvron2023llama, touvron2023llama2}, have been shown to effectively transfer knowledge obtained from web-scale data to various downstream natural language processing tasks \cite{zhao2023survey, chang2023survey}. These models are often considered backbone models due to their ability to capture rich representations of language. The Transformer architecture \cite{vaswani2017attention}, which is the foundation of many of these models, is particularly popular due to its parallelizable training paradigm.

With the recent development of the large language model, researchers started to explore recommendation language models (RLM) by reformulating various recommendation tasks into unified language tasks \cite{fan2023recommender, lin2023can, wu2023survey}. Zhang et al. \cite{zhang2021language} try to formulate user interaction history into texts and leverage the pre-trained LM to solve the sequential recommendation task, but the fine-tuned model under-performs some strong traditional baselines. Muhamed et al. \cite{muhamed2021ctr} propose CTR-BERT, which is fine-tuned from a pre-trained BERT model \cite{devlin2018bert} for the click-through-rate (CTR) prediction task. Li et al. \cite{li2023gpt4rec} adapt GPT-2 \cite{radford2019language} for title-based item sequential recommendation by using the search engine. Li et al. \cite{li2022personalized} propose PEPLER, which builds from Transformer structures for explainable recommendations with discrete and continuous prompt learning. Cui et al. \cite{cui2022m6} present M6-Rec, which builds from an existing industrial pre-trained large-scale LM by converting multiple recommendation tasks into language understanding or generation. Geng et al. \cite{geng2022recommendation} propose the P5 series, open-source fine-tuned LMs from T5 \cite{raffel2020exploring}, which can solve five different recommendation tasks in sequence-to-sequence generation form by different manually designed prompts, and outperforms many state-of-the-art recommendation models. In addition to fine-tuning the backbone large language model for recommendation tasks, another research direction involves utilizing the inherent knowledge of the backbone model through a zero-shot approach or by incorporating a few examples within the prompt as a few-shot approach \cite{sileo2022zero, bao2023tallrec, hegselmann2023tabllm, kang2023llms}.

In this paper, we apply our framework PAP-REC on P5 \cite{geng2022recommendation} as the backbone to generate automated personalized prompts since P5 is a sequence-to-sequence model with strong text generation ability on various recommendation tasks. 

\subsection{Prompt Learning}
\label{sec:prompt_learning}

The prompt is a language template containing arguments and specific trigger tokens related to the downstream tasks and can highly influence the prompt-based language model performance \cite{reynolds2021prompt}. For example, we can design a simple prompt, ``Does [user] like [item]?'' for the task of user-item preference prediction. In this example, ``[user]'' and ``[item]'' are arguments and will be filled in with actual values in the dataset. The language model receives the prompt with filled-in values as the input and generates the output sequence, expectedly ``yes'' or ``no'' for this example. However, since the language model implicitly incorporates various forms of knowledge, it is not apparent to elicit the best model performance for a given task, which motivates prompt learning.

Prompt Learning is classified into two categories based on the specific token form \cite{liu2021pre}: \textit{discrete prompt} that utilizes tokens defined in the vocabulary set, and \textit{continuous prompt} that searches the best tokens in the whole embedding space.

\subsubsection{\textbf{Discrete Prompt Learning}} Researchers have proposed many effective techniques for automated discrete prompts. Jiang et al. \cite{jiang2020can} and Yuan et al. \cite{yuan2021bartscore} update generated prompts in regards to linguistics. Gao et al. \cite{gao2021making} leverage the pre-trained T5 model \cite{raffel2020exploring} to generate discrete prompts. Deng et al. \cite{deng2022rlprompt} design a policy LM to generate prompts optimized by reinforcement learning. Wallace et al. \cite{wallace-etal-2019-universal}, and Shin et al. \cite{autoprompt:emnlp20} propose gradient-based trigger token search methods to generate prompts that lead to a stronger performance of LM. Our framework PAP-REC draws inspiration from the gradient-based methods but is capable of personalized prompt generation, differing from those of the aforementioned methods.

\subsubsection{\textbf{Continuous Prompt Learning}} Continuous embedding space allows prompts not restricted in the choice of natural language words. The larger search space contains more and potentially better prompts and, at the same time, provides a greater challenge for researchers. Li et al. \cite{li2021prefix} propose Prefix-Tuning that adds prefix vectors before encoding inputs and decoding outputs. Liu et al. \cite{liu2021gpt} design P-Tuning that inserts some learnable tokens beyond the original vocabulary set, which is essentially an automated continuous prompt. Zhong et al. \cite{zhong2021factual} leverage the discrete prompt as a starting point and optimize it in continuous embedding space. Zhou et al. \cite{zhou2022learning} apply learnable context vectors before the class name as an image-classification prompt for vision-language models. To the best of our knowledge, there are no continuous prompt learning works for RLMs. We will illustrate in Section \ref{sec:ablation_study} that directly applying existing continuous prompt learning methods can hardly generate a prompt outperforming a manual one.

\subsubsection{\textbf{Dynamic Prompt Learning}} Davison et al. \cite{davison-etal-2019-commonsense} first manually design some customized prompt examples and then leverage a unidirectional LM to select one for each input. Haviv et al. \cite{haviv2021bertese} propose BERTese for better knowledge extraction, which leverages a pre-trained LM to rewrite the input query to an input-dependence prompt. Zhang et al. \cite{zhang2022promptgen} develop an encoder-decoder model, PromptGen, to generate prompts conditional on the input sentence for knowledge probing. Zhou et al. \cite{zhou2022conditional} propose CoCoOp, an extension of CoOp \cite{zhou2022learning}, that can generate different learnable context vectors conditioning on the input images for better generalization to unseen classes. Our personalized prompts differ from the dynamic prompts since while multiple users may have distinct prompts for a single task, a single user will possess a consistent user-specific token for various queries within the same task. Besides, compared with the dynamic prompt learning methods, PAP-REC does not use manual prompts as the reference, and thus largely decreases human efforts in prompt design.

\subsubsection{\textbf{Item ID Learning}} With the recent advancements in RLMs, an associated avenue of research involves exploring the optimal item ID creation strategies tailored to diverse recommendation scenarios \cite{li2023exploring, li2023large}. Petrov et al. \cite{petrov2023generative} extract an item's ID tokens from its latent factors using Singular Value Decomposition. Hou et al. \cite{hou2023learning} employ Product Quantization (PQ) \cite{jegou2010product} to quantize item embeddings and derive their IDs. Hua et al. \cite{hua2023index} introduce three item indexing approaches—sequential indexing, collaborative indexing, and semantic indexing—conducting a comparative analysis among them. While our framework is also applicable to recommendation tasks, it differs by focusing on learning the prompt within the task description while maintaining consistent and unchanged item IDs included in \{\textit{task arg}\} in our framework. Therefore, our research is orthogonal to item ID learning, and there exists potential for a synergistic combination of these methods to complement each other in future research.

\section{Problem Formalization}
\label{sec:problem}

\begin{table}[t]
    \centering
    \begin{tabular}{c|l}
    \toprule
      {\bfseries Symbol} &{\bfseries Description}\\ 
      \midrule
      $U$ & The set of users in a recommender system\\
      $I$ & The set of items in a recommender system\\
      $R$ & The set of recommendation tasks\\
      $F$ & The recommendation language model \\
      $V$ & Vocabulary set \\
      $L$ & Loss function \\
      $M$ & Evaluation metrics \\
      $D_{train}$ & Training dataset \\
      $D_{val}$ & Validation dataset \\
      $u$ & A user ID in a recommender system\\
      $i$ & An item ID in a recommender system\\
      $[T_r]$ & A token position specific to task $r$ \\
      $[T_u]$ & A token position specific to user $u$ \\
      $t$ & A token in the vocabulary set $V$ \\
      $e_t$ & The embedding vector of token $t$ \\
      $y$ & Ground-truth token vector \\
      $y_i$ & The $i$-th token of $y$ \\
      $\hat{y}$ & Predicted token vector from the language model \\
      $x_{p}$ & A generated prompt from PAP-REC \\
      $arg$ & The argument of the prompt \\
      $r_{ui}$ & The rating of item $i$ from user $u$\\
      \bottomrule
    \end{tabular}
    \caption{Summary of the notations in this work.}
    \label{Table:notation}
\end{table}

In this section, we will first introduce the notations used in this paper (summarized in Table \ref{Table:notation}). Then, we will introduce the problem formalization separately in two parts: prompt generation and recommendation tasks. In particular, the prompt generation problem focuses on the definition of an automated prompt for the recommendation language model (RLM), while the second part discusses the downstream recommendation tasks applied in this paper. 

\begin{table*}[t]
    \centering
    \begin{tabular}{c|l|l|l}
    \toprule
      {\bfseries Task} &  {\bfseries Prompt ID} & {\bfseries Arguments} & {\bfseries Prompt}\\ 
      \midrule
      \multirow{6.5}{*}{\shortstack{Sequential\\Recommendation}} & \multirow{2}{*}{Prompt 2-3} & \multirow{2}{*}{$u$, $i$[]} & Here is the purchase history list of user\_$u$ :  \\ & & & $i$[] try to recommend next item to the user.\\
      \cmidrule{2-4}
      & \multirow{3}{*}{Prompt 2-13} & \multirow{3}{*}{$u$, $i$[]} & According to the purchase history of user\_$u$ : \\ & & &$i$[] Can you recommend the next possible\\ & & &item to the user?\\
      \midrule
      \multirow{4.5}{*}{\shortstack{Explanation\\Generation}} & \multirow{2}{*}{Prompt 3-1} & \multirow{2}{*}{$u$, title($i$)} & Generate an explanation for user\_$u$ about\\ & & & this product : title($i$) \\
      \cmidrule{2-4}
      & \multirow{2}{*}{Prompt 3-12} & \multirow{2}{*}{$u$, $i$, $r_{ui}$} & Generate a $r_{ui}$-star explanation for user\_$u$ \\ & & &about item\_$i$\\
      \midrule
      \multirow{4.5}{*}{\shortstack{User-Item\\Matching}} & \multirow{2}{*}{Prompt 5-7} & \multirow{2}{*}{$u$, $i$[]} & Pick the most suitable item from the following \\ & & & list and recommend to user\_$u$ : $i$[] \\
      \cmidrule{2-4}
      & \multirow{2}{*}{Prompt 5-8} & \multirow{2}{*}{$u$, $i$[]} & We want to make recommendation for user\_$u$. \\ & & &Select  the best item from these candidates : $i$[] \\
      \bottomrule
    \end{tabular}
    \caption{Some manual prompts used in the P5 model pre-training \cite{geng2022recommendation}.}
    \label{Table:manual}
    \vspace{-20pt}
\end{table*}

\subsection{Personalized Prompt Generation}
\label{sec:prompt_generation}

In this section, we provide a detailed introduction to the automated personalized prompt structure. 
The prompt can be seen as a function constructed by tokens and several arguments \cite{liu2021pre}. After applying the data value to the prompt arguments, we can feed the prompt sentence to the RLM and retrieve the output as the predicted value $\hat{y}$. The process can be formulated as:
\begin{equation}
  \hat{y}(arg_1, arg_2, ...) = F(x_p(arg_1, arg_2, ...))
  \label{Eq:prompt}
\end{equation}

Different downstream tasks use different tokens and arguments as prompts. Table \ref{Table:manual} displays some manually designed prompts used in the P5 model training as an example \cite{geng2022recommendation}. Since the arguments are critical task information, PAP-REC will not modify the argument contents but only the prompt sentence structure and the tokens used in the prompt sentence. Here, we define the \{\textit{task arg}\}, which contains the arguments for each task, separated by a colon. Then, the personalized prompt generated by PAP-REC can be uniformly formulated as:
\begin{equation}
  \{task\ arg\} [T_{r_1}][T_{r_2}]...[T_{r_l}][T_u]
  \label{Eq:auto prompt}
\end{equation}
where $l$ is a hyper-parameter representing the length of task-specific tokens, and $[T_u]$ is the user-specific token ($[T_u]$ will be dropped for the non-personalized prompts).
All the tokens at $[T_r]$ and $[T_u]$ are \textit{trigger tokens} as they guide the RLM to generate a proper output for the prompt sentence. Since the \textit{trigger tokens} all follow the \{\textit{task arg}\}, we also call it a suffix prompt. We will study the influence of different trigger token positions and token length $l$ in Section \ref{sec:prompt_shape}.

\subsection{Recommendation Tasks}
\label{sec:recommendation_task}

In this section, we will discuss the following three typical recommendation tasks that can be handled by the RLM: sequential recommendation, explanation generation, and user-item matching. 

\begin{itemize}
    \item \textbf{Sequential Recommendation} \cite{hidasi2015session}: Given a user ID $u$ and an item list \{$i_n$\} as the user interaction history, the RLM predicts the next item ID $i_{n+1}$ as the recommendation. In this work, the RLM chooses the whole item set $I$ as the candidate set.
    \item \textbf{Explanation Generation} \cite{zhang2020explainable}: This task requires generating an explanation sentence for a given user ID $u$ towards the given item title. To ensure a fair comparison, PAP-REC and our baselines only consider the prompt without a hint word.
    \item \textbf{User-Item Matching} \cite{rendle2009bpr}: Select the most preferred item for a given user ID $u$ from the given set of potential items. In this task, only one item in the set has interactions with the user $u$.
\end{itemize}

These three tasks can be uniformly transferred to Question-Answering (QA) tasks. Table \ref{Table:manual} shows an example of manually designed prompts as input questions for each recommendation task. Previous experiment results show that an RLM using manually designed prompts can perform better than or on par with state-of-the-art baselines on these recommendation tasks \cite{geng2022recommendation}. However, the handcrafted prompt design process requires significant human efforts to make multiple trials and hardly reaches the optimum.

\section{PAP-REC Prompt Generation Process}
\label{sec:generation_process}

\begin{figure}[t]
  \centering
  \includegraphics[width=0.8\textwidth]{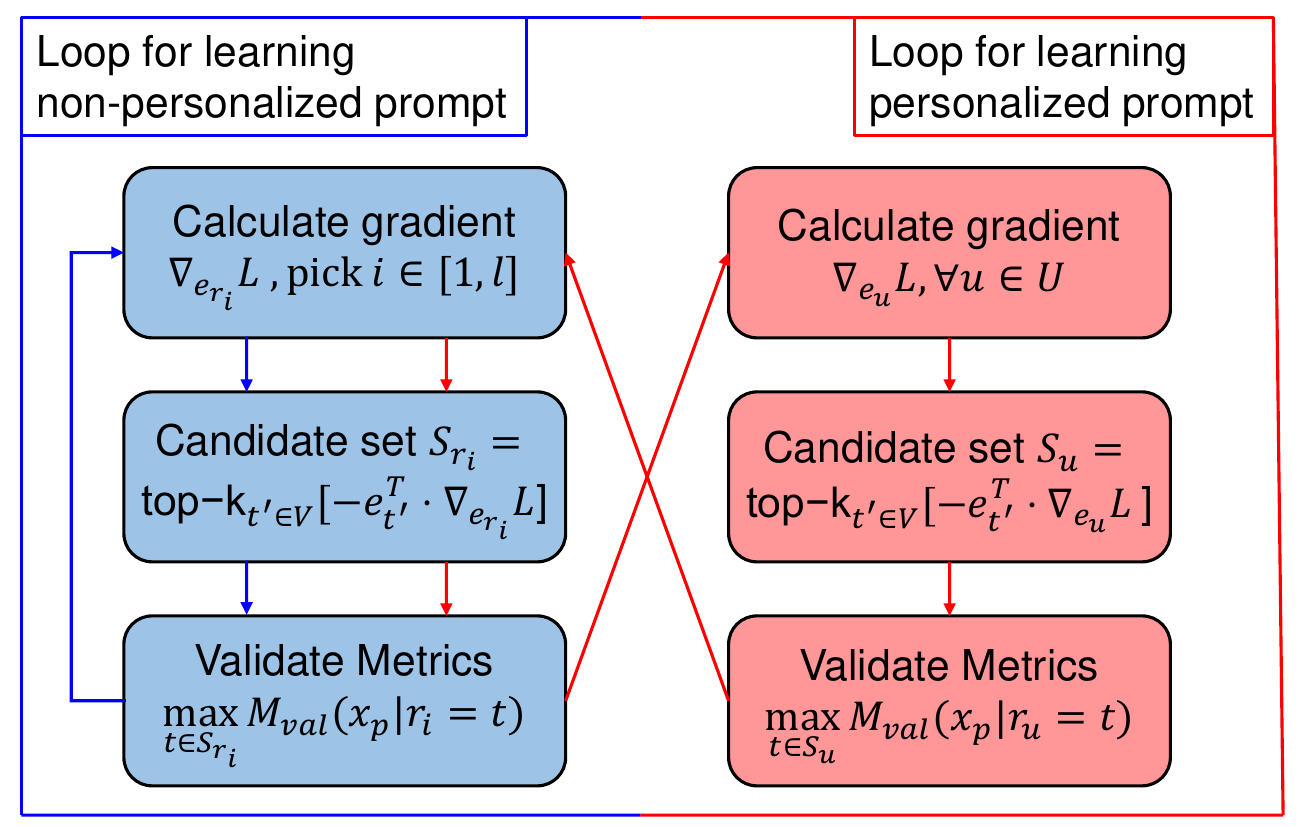}
  \caption{Overview of the iterative automated personalized and non-personalized prompt generation process.}
  \label{fig:process}
\end{figure}

In Section \ref{sec:problem}, we have introduced the automated personalized prompt and downstream recommendation tasks of PAP-REC. In this section, we will illustrate the technical details of how PAP-REC automatically generates a suitable personalized prompt for each task. Figure \ref{fig:process} provides a generation process overview. The automated non-personalized prompt generation process will follow the left blue loop in Figure \ref{fig:process}; meanwhile, the automated prompt with the user-specific token will work as the red arrows go as an alternative update schedule, which will be discussed in Section \ref{sec:alternative_personalized}. 

The subsequent sub-sections will present the challenges encountered during the personalized prompt generation process, along with the proposed solutions to address them.

\subsection{Gradient-Based Prompt Search}
\label{sec:gradient_based}

\begin{figure}[t]
  \centering
  \includegraphics[width=0.8\textwidth]{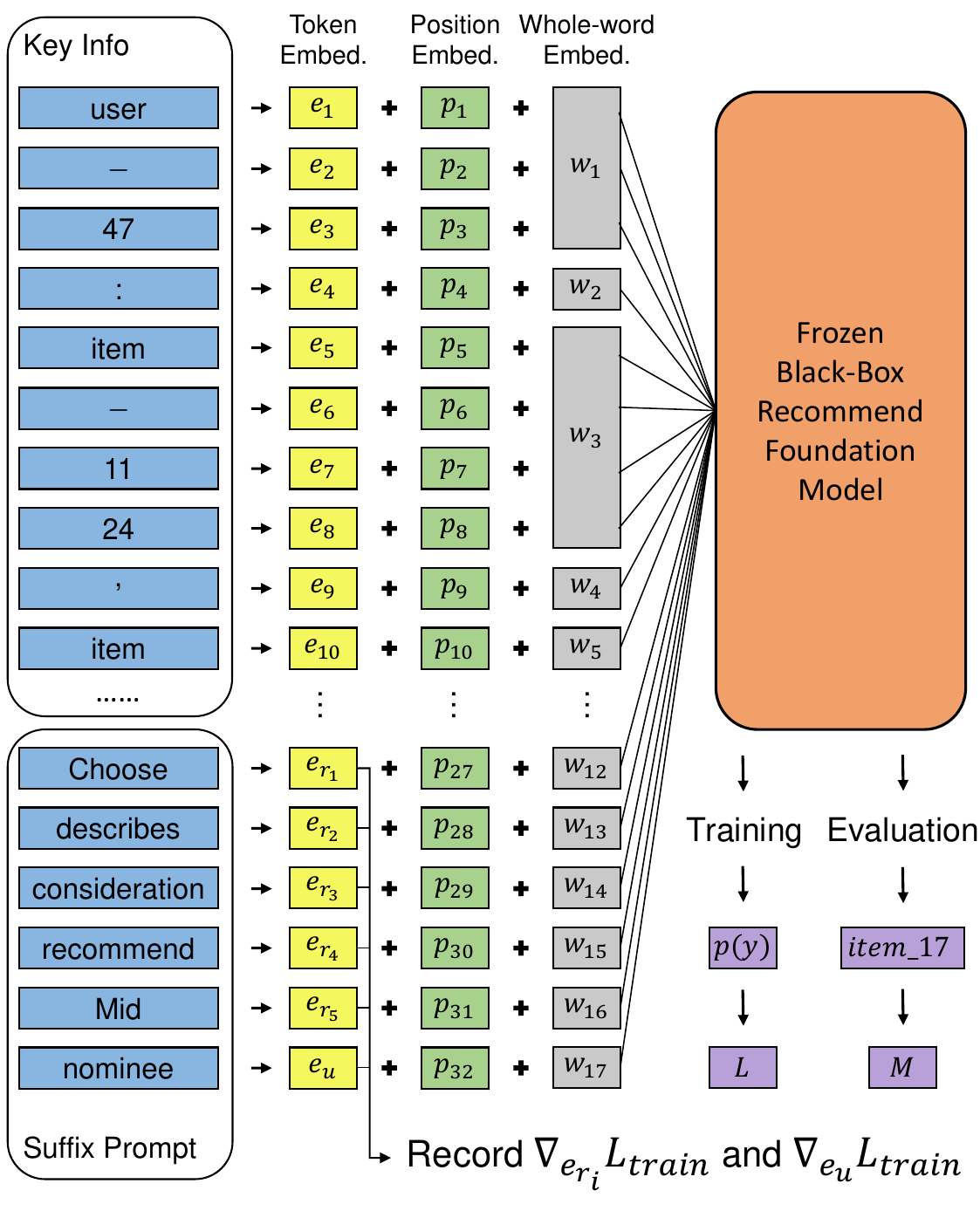}
  \vspace{-10pt}
  \caption{\small An illustration of the gradient-based prompt search method in PAP-REC. The example is an automated personalized prompt with \{\textit{task arg}\} followed by the task-specific token as ``Choose describes consideration recommend Mid'' and a specific token to the $u_{47}$, ``nominee''. PAP-REC will apply the generated prompt to the training data and record the gradient of trigger token embeddings. For non-personalized prompt search, the last token embedding $e_u$ will be $e_{r_6}$.}
  \label{fig:gradient}
\end{figure}

As mentioned in Section \ref{sec:problem}, PAP-REC needs to determine the appropriate token to fill each suffix prompt position after fixing the sentence structure. To achieve better performance, we expect to maximize the generation probability of the ground-truth value $y$ (for example, in the sequential recommendation task, $y$ is the target item ID). Thus, we define our loss function as the negative log-likelihood of the target value with the formula:
\begin{equation}
  L = -\sum_{i}^{|y|}\log p(y_i | x_p) 
  \label{Eq:loss}
\end{equation}

We encounter the first challenge as time complexity. The simplest idea to optimize our loss is to enumerate tokens in each trigger token position. However, the search time is proportional to $O(l^{|V|})$, where $|V|$ is the total number of tokens in the vocabulary set, which is unrealistic and infeasible in our work. Thus, we need a more efficient algorithm to shrink the token candidate sets. We propose our method inspired by the recent works on gradient-based prompt search \cite{wallace-etal-2019-universal, autoprompt:emnlp20}. Figure \ref{fig:gradient} provides an illustration of our method with a generated prompt as an example, and we illustrate the theory below.

The loss function after changing the trigger position $[T_{r_i}]$ from $t$ to $t'$ can be approximated with the first-order polynomial as:
\begin{equation}
  L([T_{r_i}] = t') \approx L([T_{r_i}] = t) + (e_{t'} - e_{t})^T \cdot \nabla_{e_{r_i}} L([T_{r_i}] = t)
  \label{Eq:approximation}
\end{equation}
where $e_{r_i}$ is the $i$-th task-specific token embedding. With the updated token $t'$ as the variable, the only non-constant term of Eq.\eqref{Eq:approximation} is:
\begin{equation}
  e_{t'}^T \cdot \nabla_{e_{r_i}} L([T_{r_i}] = t)
  \label{Eq:non-constant}
\end{equation}

Therefore, based on our loss decrement goal, we define the candidate token set $S_{r_i}$ as:
\begin{equation}
  S_{r_i} = \mathop{\text{top-k}}\limits_{t'\in V}\big[-e_{t'}^T \cdot \nabla_{e_{r_i}} L([T_{r_i}] = t)\big]
  \label{Eq:candidate_set}
\end{equation}

We select top-k tokens as candidates instead of directly using the top-1 because of the linear approximation applied to the loss function, and the token choices are expected to converge after several rounds of replacement. We only convert one token position each round in practice, as replacing one token will change the gradient of other $[T_r]$. PAP-REC randomly selects one $[T_{r_i}]$ to replace in each round to avoid getting stuck at the local optimum. 

Besides, PAP-REC can generate automated personalized prompts for any sequence-to-sequence recommendation language model (RLM) since the RLM is frozen as a black box in Figure \ref{fig:gradient}.

\subsection{Surrogate Recommendation Metrics}
\label{sec:surrogate_metrics}

In order to efficiently minimize the training loss through the generated prompt, we have built the candidate token set $S_{r_i}$ at each position based on Eq.\eqref{Eq:candidate_set}. A new problem derived from the candidate set is the criterion for selecting the most appropriate token replacement among the top-k candidates. Some existing works leverage the validation dataset or a pre-reserved part of the training dataset to evaluate the loss decrement after token replacement and select the token with the largest loss decrement \cite{wallace-etal-2019-universal, autoprompt:emnlp20}.

This method may be effective for typical Natural Language Processing (NLP) problems such as sentiment analysis and natural language inference; however, we will show in the experiment of Section \ref{sec:ablation_study} that recommendation metrics could not benefit from using the loss as the selection criterion for the following reasons:

First, most recommendation metrics are non-differentiable, and researchers need to design some surrogate losses to approximate the optimal goal. Thus, minimizing the negative log-likelihood loss can only improve the recommendation performance to some extent. Second, unlike typical NLP problems, recommender systems usually focus on top-k outputs instead of the most likely single one when evaluating under metrics such as Hit Rate (HR) and Normalized Discounted Cumulative Gain (NDCG). Maximizing the probability of generating target tokens may not lead to the generalization of the top-k evaluation metrics. Out of such consideration, we need to design a better selection criterion specific to the recommendation tasks.

Noticing that the selection criterion does not need to be differentiable, we decide to use the evaluation metrics on the validation dataset to select the best token replacement. The idea is consistent with the reward of reinforcement learning methods \cite{deng2022rlprompt} and some automated machine learning works \cite{pham2018efficient, li2022autolossgen}. In this way, our optimization problem can be formulated as:
\begin{equation}
\begin{split}
  \max\limits_{t\in S_{r_i}} \text{ } & M_{val}\big(x_p | [T_{r_i}] = t\big) \\
  \text{s.t.~~~} & \{S_{r_i}\} = \mathop{\text{top-k}}\limits_{t'\in V} \Big[\text{arg}\min L_{train}\big(x_p|[T_{r_i}] = t'\big)\Big]
\end{split}
\label{Eq:optimization}
\end{equation}

Although leveraging evaluation metrics as the selection criterion solves the problem derived from top-k evaluation metrics, the evaluation process is too time-consuming. Table \ref{Table:time comparison} displays an example of token evaluation running time. Since the token update process is iterative, the whole evaluation time must multiply the number of candidate tokens $|S_{r_i}|$ and the number of convergence epochs. Over ten times difference between $L$ and $M$ on running efficiency in Table \ref{Table:time comparison} could increase the search process from about one day to several days or even weeks on larger models or datasets, which is impracticable. To address this problem, we design the surrogate metrics $\widetilde{M}$ for approximation. $\widetilde{M}$ improves the efficiency in the following aspects:

\begin{itemize}
    \item The language model employs beam search to generate multiple outputs with the highest probability. We set the beam size as twenty in the final test to encourage a thorough output search, while $\widetilde{M}$ decreases the beam size to five for speed-up, though some higher probability outputs may be dropped.
    \item $\widetilde{M}$ only considers the weighted sum of part of the evaluation metrics. Take the sequential recommendation task as an example; the language model needs a beam size of at least ten to generate ten different outputs to evaluate metric@10 in the final test, while the surrogate metrics limit the size of beam search and the output as five and assign $\widetilde{M} = HR@5 + \alpha_{seq} \cdot NDCG@5$, where the hyper-parameter $\alpha_{seq}$ is a balancing coefficient of the sequential recommendation task.
\end{itemize}

\begin{table}[t]
    \centering
    \begin{tabular}{l|r}
    \toprule
      {\bfseries Selection Criterion} &{\bfseries Running time}\\ 
      \midrule
      Loss function $L$ & $\sim$1.5\text{ }minutes \\
      Evaluation metrics $M$ & $\sim$25\text{ }minutes \\
      Surrogate metrics $\widetilde{M}$ & $\sim$4\text{ }minutes \\
      \bottomrule
    \end{tabular}
    \caption{Time comparison between different selection criteria. We use the P5-S as the backbone recommendation language model and \textit{Beauty} as the dataset to evaluate one token replacement under the same environment.}
    \label{Table:time comparison}
\end{table}

According to the descriptions above, the surrogate metrics $\widetilde{M}$ sacrifice some accuracy for better efficiency. As Table \ref{Table:time comparison} shows, the surrogate metrics bring about six times speed-up than $M$; and using $\widetilde{M}$ is only about twice the prompt generation time than using $L$ because the token evaluation only takes part of the generation process. We will show in the experiment that using surrogate metrics can generate prompts outperforming the manual ones.

\subsection{Alternative Personalized Prompt Update}
\label{sec:alternative_personalized}

Personalization is one of the essential features of recommendation systems (RS). In this work, we propose the concept of an automated personalized prompt by introducing an extra user-specific token position $[T_u]$ after the task-specific tokens $\{[T_{r_i}]\}$. Similar to our personalized prompts, in the NLP field, dynamic prompts offer a customized template for each input \cite{liu2021pre}; 
however, our personalized prompts differ from the dynamic prompts since while multiple users may have distinct prompts for a single task, a single user will possess a consistent user-specific token for various queries within the same task.

Leveraging the equations that we derived in previous sub-sections, we can define the candidate set $S_u$ for $[T_u]$ in the same way as Eq.\eqref{Eq:candidate_set}:
\begin{equation}
  \forall u\in U, \text{ } S_{u} = \mathop{\text{top-k}}\limits_{t'\in V}\big[-e_{t'}^T \cdot \nabla_{e_{u}} L([T_{u}] = t)\big]
  \label{Eq:user_candidate_set}
\end{equation}
where $e_u$ is the user-specific token embedding. Then, we have a similar optimization problem to Eq.\eqref{Eq:optimization}:
\begin{equation}
\begin{split}
  \forall u\in U, \text{ } & \max\limits_{t\in S_{u}} \text{ } M_{val}\big(x_p | [T_{u}] = t\big) \\
  \text{s.t.~~~} & \{S_{u}\} = \mathop{\text{top-k}}\limits_{t'\in V} \Big[\text{arg}\min L_{train}\big(x_p|[T_{u}] = t'\big)\Big]
\end{split}
\label{Eq:user_optimization}
\end{equation}

However, a new challenge emerges as the inflating search space. For an RS with 30K different user IDs, if we preset $|\{[T_{r_i}]\}| = 5$ and $|\{[T_u]\}| = 1$ for each user, then a personalized prompt will have six thousand more searchable token positions than a non-personalized prompt. If PAP-REC updates $[T_u]$ with the same schedule as $[T_r]$, it is infeasible to generate an effective personalized prompt in a practical time. 

To address the inflating search space problem, we notice that for all three recommendation tasks in our work, each piece of data will only contain one user ID, so there are no overlaps between the data of different users. In other words, if we denote the dataset related to the user $u$ by $D(u)$, we have:
\begin{equation}
\begin{split}
  & \forall i, j \in U, \text{ } i \neq j \iff D(i) \cap D(j) = \emptyset \\
  & \Rightarrow \sum_{u\in U} D_{train}(u) = D_{train} \text{ and } \sum_{u\in U} D_{val}(u) = D_{val}
\end{split}
\label{Eq:user_intersection}
\end{equation}
and the optimization problem of Eq.\eqref{Eq:user_optimization} can be reformulated as:
\begin{equation}
\begin{split}
  \forall u\in U, \text{ } & \max\limits_{t\in S_{u}} \text{ } M_{D_{val}(u)}\big(x_p | [T_{u}] = t\big) \\
  \text{s.t.~~~} & \{S_{u}\} = \mathop{\text{top-k}}\limits_{t'\in V} \Big[\text{arg}\min L_{D_{train}(u)}\big(x_p|[T_{u}] = t'\big)\Big]
\end{split}
\label{Eq:user_optimization_reformulate}
\end{equation}

In this way, when updating the user-specific token $[T_u]$, we only leverage $D_{train}(u)$ to calculate the gradient of $[T_u]$ and $D_{val}(u)$ to validate the metrics of token replacement. The sum of the running time is compatible with updating the task-specific token.

Another potential challenge from the personalized prompt is the scarcity of data within $D(u)$ after allocating the complete dataset to different users. This situation raises the issue of scarce data when updating the user-specific tokens. Recognizing the significance of addressing the scarce data problem, we introduce an iterative and alternative update schedule for $[T_u]$ to effectively resolve this concern. 

First, to avoid mutual influence, $[T_u]$'s update will be in a different epoch from that of $[T_r]$. We expect that $[T_u]$ and $[T_r]$ will converge to the optimal after multiple iterations of updating. Second, we do not separate the training data into batches due to the sparsity and limited amount of $D(u)$ for each user. 

Finally, the personalized prompt learning loop is iterative and alternative, as the red arrows go in Figure \ref{fig:process}. In the test phase, we will use the validation dataset to evaluate generated prompts after every token update and select the prompt with the best overall performance on various metrics. Then, the selected prompt is used on the test dataset for the final performance evaluation.

\begin{table}[t]
    \centering
    \begin{tabular}{c|cccc}
        \toprule
         Dataset & \#Users & \#Items & \#Interactions & Density\\
         \midrule
         Beauty & 22,363 & 12,101 & 198,502 & 0.0734\%\\
         Sports & 35,598 & 18,357 & 296,337 & 0.0453\%\\
         Toys & 19,412 & 11,924 & 167,597 & 0.0724\%\\
         \bottomrule
    \end{tabular}
    \caption{Basic statistics of the datasets.}
    \label{Table:dataset}
\end{table}

\begin{sidewaystable}[htp]
    \centering
    \begin{tabular}{c|c|c|l}
    \toprule
         Recommendation Task & Dataset & Backbone & Decoding Prompt \\
         \midrule
         \multirow{7}{*}{\shortstack{Sequential\\Recommendation}} & \multirow{2}{*}{Beauty} & P5-S & user\_$u$: $i[]$ Choose describes consideration recommend Mid nominee \\
         & & P5-B & user\_$u$: $i[]$ Imagine item item cumpara next \\
         \cmidrule{2-4}
         & \multirow{2}{*}{Sports} & P5-S & user\_$u$: $i[]$ Choose contingent next discuss Users \\
         & & P5-B & user\_$u$: $i[]$ Imagine allergy next item next \\
         \cmidrule{2-4}
         & \multirow{2}{*}{Toys} & P5-S & user\_$u$: $i[]$ recommend describe list Possible suprafete \\
         & & P5-B & user\_$u$: $i[]$ Choose omul next consumer next \\
         \midrule
         \multirow{7}{*}{\shortstack{Explanation\\Generation}} & \multirow{2}{*}{Beauty} & P5-S & user\_$u$: title($i$) Cur fisier explanation Desktop compoziti \\
         & & P5-B & user\_$u$: title($i$) explanation allergy Gene achizitiona gasest \\
         \cmidrule{2-4}
         & \multirow{2}{*}{Sports} & P5-S & user\_$u$: title($i$) machiaj Elizabeth explanation Converter apa \\
         & & P5-B & user\_$u$: title($i$) Give with word caption explanation \\
         \cmidrule{2-4}
         & \multirow{2}{*}{Toys} & P5-S & user\_$u$: title($i$) disappointment Elizabeth 2) explanation helps \\
         & & P5-B & user\_$u$: title($i$) Potential explanation explanation Duration disappoint \\
         \midrule
         \multirow{7}{*}{\shortstack{User-Item\\Matching}} & \multirow{2}{*}{Beauty} & P5-S & user\_$u$: $i[]$ 1910 Elizabeth gradina gradina 1800 \\
         & & P5-B & user\_$u$: $i[]$ fancy guides 65 API gradini \\
         \cmidrule{2-4}
         & \multirow{2}{*}{Sports} & P5-S & user\_$u$: $i[]$ pasageri describe românesc educație Simmons \\
         & & P5-B & user\_$u$: $i[]$ emb guides älter API fla \\
         \cmidrule{2-4}
         & \multirow{2}{*}{Toys} & P5-S & user\_$u$: $i[]$ Gratuit Elizabeth negru snowboard urmari \\
         & & P5-B & user\_$u$: $i[]$ trop guides accelerate 1906 proaspat \\
         \bottomrule
    \end{tabular}
    \vspace{5pt}
    \caption{Generated automated non-personalized prompts from PAP-REC.}
    \label{Table:auto_prompt}
    \vspace{-20pt}
\end{sidewaystable}

\section{Experiments}
\label{sec:experiment}

In this section, we conduct experiments to provide a better understanding of the personalized prompt generation process and validate the effectiveness of automatically generated prompts by the proposed framework---PAP-REC\footnote{The source code of the work is available at https://github.com/rutgerswiselab/PAP-REC.}.

\subsection{Experimental Setup}
\label{sec:experimental_setup}

\subsubsection{\textbf{Recommendation Datasets and Tasks}}
\label{sec:dataset}

We conduct our experiments on three benchmark datasets of recommendation systems (RS) from \textit{Amazon.com}, namely \textit{Sports}, \textit{Beauty}, and \textit{Toys}, which are publicly available\footnote{\url{https://huggingface.co/datasets/amazon_us_reviews}}. We follow the existing works \cite{zhou2020s3, geng2022recommendation} to generate the dataset spanning from January 1, 2019, to December 31, 2019. The detailed statistics of the datasets are shown in Table \ref{Table:dataset}.

In this work, we research automated personalized prompts on three recommendation tasks as mentioned in Section \ref{sec:recommendation_task}: sequential recommendation, explanation generation, and user-item matching. For both sequential recommendation and user-item matching, we select the last item of each user as the ground truth in the test dataset, the second-to-last item in the validation dataset, and the remaining items in the training dataset, based on the timestamp of the user interaction history. For the sequential recommendation, upon selecting the ground truth $y$, we restrict the interactions that occurred prior to $y$ as part of \{\textit{task arg}\} in the prompt to avoid data leakage. For the user-item matching task, we randomly select at most 99 items outside the user purchase history, along with $y$, to build the candidate item set for the target item uniqueness. To construct the explanation dataset, we adapt the methodology employed in previous studies \cite{li2020generate, li2021personalized} that focus on natural language explanations. Initially, we utilize the Sentires toolkit3 \cite{zhang2014explicit, zhang2014users} to extract item feature words under consideration from customer reviews. Subsequently, we extract sentences from these reviews that comment on one or more item feature words as users’ explanations about their preferences.

\subsubsection{\textbf{Evaluation Metrics}}
\label{sec:metrics}

To evaluate explanation generation, we use typical metrics BLEU-4 \cite{papineni2002bleu}, as well as ROUGE-1, ROUGE-2, and ROUGE-L \cite{lin2004rouge} for comparison between the ground truth and the generated output from the recommendation language model (RLM). For the other recommendation tasks, we use the RLM to generate top-10 different outputs with the highest probability and apply Hit Ratio (HR) @5, @10 \cite{burke2005segment} and Normalized Discounted Cumulative Gain (NDCG) @5, @10 \cite{jarvelin2002cumulated, buttcher2016information}, as evaluation metrics. Besides, we also report the HR@1 for the matching task due to the uniqueness of the target item. 
All metrics are the higher, the better.

\subsubsection{\textbf{Language Models}}
\label{sec:model}

We apply our PAP-REC to two versions of P5 \cite{geng2022recommendation} for personalized prompt generation, namely P5-small (P5-S) and P5-base (P5-B). As for the differences between these two model versions, P5-B has a larger embedding dimension and more layers of Transformer blocks. P5 is a very suitable RLM for our PAP-REC as it is a sequence-to-sequence model, and P5 with manual prompts outperforms state-of-the-art non-LLM recommendation models on multiple tasks with manual prompts in the experiment of \cite{geng2022recommendation}. We will show that automated personalized prompts generated from PAP-REC enable P5 to enhance the recommendation performance further. 

\subsubsection{\textbf{\textbf{Baselines}}}
\label{sec:baselines}

As PAP-REC generates prompts for the RFM, the outputs of the baselines should also be prompts. P5, as a prompt-based model, provides several manually designed prompts in \cite{geng2022recommendation} for the three tasks. We leverage the prompts containing the same arguments with our automated prompt in each task as manual prompt sets and select the one with the best performance from the set as the baseline. Specifically, as shown in Table \ref{Table:manual}, we use Prompt 2-3 for the sequential recommendation, Prompt 3-1 for the explanation generation, and Prompt 5-7 for the user-item matching as the baseline, namely \textit{Seen manual} prompt. Besides, we also leverage \textit{Unseen manual} prompts that are mentioned in \cite{geng2022recommendation} but not used in the P5 pre-training process as another baseline, 
i.e., Prompt 2-13\footnote{\label{unseen}The manual prompt can be found in the Appendix of \cite{geng2022recommendation}.} for the sequential task, Prompt 3-12\footref{unseen} for the explanation generation, and Prompt 5-8\footref{unseen} for the matching task. 

\subsubsection{\textbf{Implementation Details}}
\label{sec:implementation}

Our framework is implemented by PyTorch, an open-source library. As we mentioned in Section \ref{sec:prompt_generation}, PAP-REC can generate two types of automated prompts: non-personalized prompts and personalized prompts. 
We use suffix prompts as default, i.e., all trigger words follow the \{\textit{task arg}\}. We set the length of task-specific tokens for personalized prompts $l$ as 5, followed by only one single user-specific token $[T_u]$ due to the sparsity of interactions per user, as shown in Table \ref{Table:dataset}. Since personalized prompts have 6 tokens in total, including 5 tokens specific to the task, for a fair comparison, we set the length of non-personalized prompts $|\{[T_r]\}|$ as 5 or 6 and select the better one based on the validation set. The \{\textit{task arg}\} contains a user ID and an item title for the explanation generation, a user ID and an item ID list as the interaction history for the sequential recommendation, and a user ID and an ID list of item candidates for the user-item matching. 

During the automated prompt generation process, PAP-REC leverages the negative log-likelihood loss for gradient calculation and selects top-k candidate tokens for each token position. We set the top-k as top-5 as default. As mentioned in Section \ref{sec:surrogate_metrics}, we assign surrogate metrics $\widetilde{M} = HR@5 + \alpha_{seq} \cdot NDCG@5$ for sequential recommendation and user-item matching tasks, 
while for explanation generation, we assign $\widetilde{M} = \text{BLEU\_4} + \alpha_{exp} \cdot \text{ROUGE\_L}$. We set the balanced coefficients $\alpha_{seq}=1$ and $\alpha_{exp}=0.1$ for the best practice. Besides, for the personalized prompt search, the user-specific token $[T_u]$ is updated in a separated and alternative epoch with $[T_r]$, and we set $[T_u] =$ "?" as default to avoid the exception of some missing user IDs in the training and validation datasets. We set the maximum number of updating epochs as 50 for both types of prompts.

Finally, when testing the prompt performance, we set the beam size of P5 output generation as 20 to evaluate prompt performance. For each model on each dataset, we run the testing process 5 times and report the average result.

\subsection{Generated Prompts}
\label{sec:generated_prompt}

In this section, we show the automated prompts under the generation process of our framework described in Section \ref{sec:generation_process} and Figure \ref{fig:process}. We run the PAP-REC framework under eighteen task-dataset-model combinations. 
We only display the best generated non-personalized prompt under each dataset-task-model combination in Table \ref{Table:auto_prompt} because personalized prompts are different for every user.

We can see that all the trigger tokens follow the \{\textit{task arg}\}, which varies depending on the recommendation tasks in Table \ref{Table:auto_prompt}. The sentence constructed by trigger tokens is not meaningful from human understanding, which is consistent with some existing works \cite{autoprompt:emnlp20, zhang2022promptgen}. Although the trigger tokens are not immediately readable, we can still have some observations on the generated tokens. For each recommendation task, there are some high-frequency tokens related to the task. For example, in the sequential task, ``recommend'', ``next'', ``item'', and ``choose'' are generated many times, and ``explanation'' is the predominant keyword for the explanation generation task, which is consistent with human intuition. The other tokens can be seen as the ``auxiliary'' tokens to boost the backbone for better performance. 

For the user-item matching task, however, we can hardly find some co-occurring tokens related to the task. A possible reason is that the P5 model does not see the target item when training but maybe some negative samples related to the user ID. Thus, it is easier to recognize the negative samples than predict the next item with the given purchase history, though the matching task and the sequential task have similar forms of \{\textit{task arg}\}. With a relatively easy task, the trigger words do not need to be extremely powerful to supervise the model but provide some auxiliary guidance to lead to the emergence of strong performance.

\begin{table}[t]
    \centering
    \setlength{\tabcolsep}{2pt}
    \begin{tabular}{c|c|cccc}
        \toprule
        \multirow{2.5}{*}{Dataset} & \multirow{2.5}{*}{Metrics} & \multicolumn{4}{c}{Prompts} \\
        \cmidrule{3-6}
        & & Seen manual & Unseen manual & Non-personalized (Our) & Personalized (Our)  \\
        \midrule
        \multirow{4}{*}{Beauty} & HR@5 & 0.0536 & 0.0524 & 0.0546 & \bm{$0.0559^*$} \\
        & NDCG@5 & 0.0401 & 0.0389 & 0.0408 & \bm{$0.0414^*$} \\
        & HR@10 & 0.0665 & 0.0640 & 0.0682 & \bm{$0.0688^*$} \\
        & NDCG@10 & 0.0443 & 0.0426 & 0.0452 & \bm{$0.0456^*$} \\
        \midrule
        \multirow{4}{*}{Sports} & HR@5 & 0.0402 & 0.0386 & 0.0406 & \bm{$0.0408$} \\
        & NDCG@5 & 0.0330 & 0.0315 & \bm{$0.0331$} & 0.0329 \\
        & HR@10 & 0.0462 & 0.0442 & \bm{$0.0469$} & \bm{$0.0469$} \\
        & NDCG@10 & 0.0349 & 0.0333 & \bm{$0.0351$} & 0.0349 \\
        \midrule
        \multirow{4}{*}{Toys} & HR@5 & 0.0632 & 0.0632 & \bm{$0.0650$} & 0.0644 \\
        & NDCG@5 & 0.0550 & 0.0546 & \bm{$0.0562^*$} & 0.0555 \\
        & HR@10 & 0.0711 & 0.0704 & \bm{$0.0737^*$} & 0.0729 \\
        & NDCG@10 & 0.0576 & 0.0569 & \bm{$0.0591^*$} & 0.0582 \\
        \bottomrule
    \end{tabular}
    \caption{Performance on sequential recommendation task using P5-S as the backbone. Bold numbers represent the best performance of each row. * indicates its performance is significantly better at $\bm{p < 0.05}$ than all other prompts.}
    \label{Table:sequential_S}
\end{table}

\begin{table}[t]
\setlength{\tabcolsep}{2pt}
    \centering
    \begin{tabular}{c|c|cccc}
        \toprule
        \multirow{2.5}{*}{Dataset} & \multirow{2.5}{*}{Metrics} & \multicolumn{4}{c}{Prompts} \\
        \cmidrule{3-6}
        & & Seen manual & Unseen manual & Non-personalized (Our) & Personalized (Our)  \\
        \midrule
        \multirow{4}{*}{Beauty} & HR@5 & 0.0476 & 0.0468 & \bm{$0.0478$} & \bm{$0.0478$} \\
        & NDCG@5 & 0.0359 & 0.0354 & 0.0359 & \bm{$0.0360$} \\
        & HR@10 & 0.0596 & 0.0591 & 0.0600 & \bm{$0.0602$} \\
        & NDCG@10 & 0.0398 & 0.0394 & 0.0398 & \bm{$0.0400$} \\
        \midrule
        \multirow{4}{*}{Sports} & HR@5 & 0.0369 & 0.0362 & \bm{$0.0391$} & 0.0389\\
        & NDCG@5 & 0.0307 & 0.0301 & \bm{$0.0327$} & 0.0322 \\
        & HR@10 & 0.0437 & 0.0427 & \bm{$0.0465$} & 0.0459 \\
        & NDCG@10 & 0.0329 & 0.0322 & \bm{$0.0352^*$} & 0.0345 \\
        \midrule
        \multirow{4}{*}{Toys} & HR@5 & 0.0578 & 0.0567 & \bm{$0.0588$} & 0.0585 \\
        & NDCG@5 & 0.0485 & 0.0465 & 0.0490 & \bm{$0.0492$} \\
        & HR@10 & 0.0656 & 0.0639 & 0.0666 & \bm{$0.0671$} \\
        & NDCG@10 & 0.0510 & 0.0488 & 0.0515 & \bm{$0.0520^*$} \\
        \bottomrule
    \end{tabular}
    \caption{Performance on sequential recommendation task using P5-B as the backbone. Bold numbers represent the best performance of each row. * indicates its performance is significantly better at $\bm{p < 0.05}$ than all other prompts.}
    \label{Table:sequential_B}
\end{table}

\begin{table}[t]
\setlength{\tabcolsep}{2pt}
    \centering
    \begin{tabular}{c|c|cccc}
        \toprule
        \multirow{2.5}{*}{Dataset} & \multirow{2.5}{*}{Metrics} & \multicolumn{4}{c}{Prompts} \\
        \cmidrule{3-6}
        & & Seen manual & Unseen manual & Non-personalized (Our) & Personalized (Our)  \\
        \midrule
        \multirow{4}{*}{Beauty} & BLEU-4 & 1.2153 & 1.1332 & \bm{$1.2540$} & 1.2418 \\
        & ROUGE-1 & 17.7581 & 16.2971 & \bm{$17.9697$} & 17.9029 \\
        & ROUGE-2 & 2.2411 & 1.9700 & \bm{$2.3659$} & 2.3401 \\
        & ROUGE-L & 12.9374 & 12.0170 & \bm{$13.0855$} & 13.0408 \\
        \midrule
        \multirow{4}{*}{Sports} & BLEU-4 & 0.5545 & 0.5786 & \bm{$0.6037^*$} & 0.5720 \\
        & ROUGE-1 & 13.2365 & 11.6826 & \bm{$13.7212$} & 13.5409 \\
        & ROUGE-2 & 1.3168 & 1.2125 & 1.3168 & \bm{$1.3431^*$} \\
        & ROUGE-L & 9.7524 & 9.0215 & 9.9198 & \bm{$9.9218$} \\
        \midrule
        \multirow{4}{*}{Toys} & BLEU-4 & 1.5714 & 1.6525 & 1.8435 & \bm{$1.9584^*$} \\
        & ROUGE-1 & 13.4572 & 13.6342 & 14.5658 & \bm{$14.7814^*$} \\
        & ROUGE-2 & 3.3849 & 3.3739 & 3.4539 & \bm{$3.6412^*$} \\
        & ROUGE-L & 10.7131 & 10.7361 & 11.3594 & \bm{$11.5336^*$} \\
        \bottomrule
    \end{tabular}
    \caption{Performance on explanation generation task using P5-S as the backbone (all numbers in this table are percentage numbers with "\%" omitted for clarity, e.g., 1.2153 means 1.2153\%). Bold numbers represent the best performance of each row. * indicates its performance is significantly better at $\bm{p < 0.05}$ than all other prompts.}
    \label{Table:explanation_S}
\end{table}

\begin{table}[t]
\setlength{\tabcolsep}{2pt}
    \centering
    \begin{tabular}{c|c|cccc}
        \toprule
        \multirow{2.5}{*}{Dataset} & \multirow{2.5}{*}{Metrics} & \multicolumn{4}{c}{Prompts} \\
        \cmidrule{3-6}
        & & Seen manual & Unseen manual & Non-personalized (Our) & Personalized (Our)  \\
        \midrule
        \multirow{4}{*}{Beauty} & BLEU-4 & 0.8793 & 0.6773 & \bm{$0.9407^*$} & 0.9272 \\
        & ROUGE-1 & \bm{$16.3777$} & 14.1965 & 16.2540 & 16.1988 \\
        & ROUGE-2 & 1.7706 & 1.3030 & \bm{$1.7824$} & 1.7761 \\
        & ROUGE-L & \bm{$11.7751$} & 10.3834 & 11.6643 & 11.6130 \\
        \midrule
        \multirow{4}{*}{Sports} & BLEU-4 & 0.5665 & 0.2491 & \bm{$0.6159^*$} & 0.5924 \\
        & ROUGE-1 & 12.5667 & 11.0995 & \bm{$13.1524^*$} & 12.8222 \\
        & ROUGE-2 & 1.1661 & 0.6779 & \bm{$1.3462^*$} & 1.2634 \\
        & ROUGE-L & 9.4110 & 8.4553 & \bm{$9.7217^*$} & 9.5452 \\
        \midrule
        \multirow{4}{*}{Toys} & BLEU-4 & 1.8907 & 1.5242 & \bm{$2.0366^*$} & 1.9255 \\
        & ROUGE-1 & 14.5309 & 12.6479 & 14.1106 & \bm{$14.6687^*$} \\
        & ROUGE-2 & 3.5776 & 3.0812 & \bm{$3.7356^*$} & 3.5490 \\
        & ROUGE-L & 11.5472 & 10.2753 & 11.2934 & \bm{$11.5669$} \\
        \bottomrule
    \end{tabular}
    \caption{Performance on explanation generation task using P5-B as the backbone (all numbers in this table are percentage numbers with "\%" omitted for clarity, e.g., 0.8793 means 0.8793\%). Bold numbers represent the best performance of each row. * indicates its performance is significantly better at $\bm{p < 0.05}$ than all other prompts.}
    \label{Table:explanation_B}
\end{table}

\begin{table}[t]
\setlength{\tabcolsep}{2pt}
    \centering
    \begin{tabular}{c|c|cccc}
        \toprule
        \multirow{2.5}{*}{Dataset} & \multirow{2.5}{*}{Metrics} & \multicolumn{4}{c}{Prompts} \\
        \cmidrule{3-6}
        & & Seen manual & Unseen manual & Non-personalized (Our) & Personalized (Our)  \\
        \midrule
        \multirow{5}{*}{Beauty} & HR@1 & 0.0587 & \bm{$0.0633^*$} & 0.0620 & 0.0613 \\
        & HR@5 & 0.1583 & 0.1573 & 0.1608 & \bm{$0.1616$} \\
        & NDCG@5 & 0.1094 & 0.1112 & 0.1125 & \bm{$0.1126$} \\
        & HR@10 & 0.2338 & 0.2283 & \bm{$0.2377$} & 0.2369 \\
        & NDCG@10 & 0.1337 & 0.1340 & \bm{$0.1372$} & 0.1367 \\
        \midrule
        \multirow{5}{*}{Sports} & HR@1 & 0.0669 & 0.0660 & \bm{$0.0673$} & 0.0662 \\
        & HR@5 & 0.1813 & 0.1812 & \bm{$0.1831$} & 0.1821 \\
        & NDCG@5 & 0.1253 & 0.1245 & \bm{$0.1262$} & 0.1253 \\
        & HR@10 & 0.2666 & 0.2636 & \bm{$0.2667$} & 0.2656 \\
        & NDCG@10 & 0.1526 & 0.1509 & \bm{$0.1531$} & 0.1521 \\
        \midrule
        \multirow{5}{*}{Toys} & HR@1 & 0.0434 & 0.0423 & 0.0438 & \bm{$0.0448^*$} \\
        & HR@5 & 0.1265 & 0.1283 & 0.1287 & \bm{$0.1304^*$} \\
        & NDCG@5 & 0.0852 & 0.0860 & 0.0862 & \bm{$0.0878^*$} \\
        & HR@10 & 0.2016 & 0.1991 & \bm{$0.2038$} & \bm{$0.2038$} \\
        & NDCG@10 & 0.1092 & 0.1087 & 0.1103 & \bm{$0.1113^*$} \\
        \bottomrule
    \end{tabular}
    \caption{Performance on user-item matching task using P5-S as the backbone. Bold numbers represent the best performance of each row. * indicates its performance is significantly better at $\bm{p < 0.05}$ than all other prompts.}
    \label{Table:matching_S}
\end{table}

\begin{table}[t]
\setlength{\tabcolsep}{2pt}
    \centering
    \begin{tabular}{c|c|cccc}
        \toprule
        \multirow{2.5}{*}{Dataset} & \multirow{2.5}{*}{Metrics} & \multicolumn{4}{c}{Prompts} \\
        \cmidrule{3-6}
        & & Seen manual & Unseen manual & Non-personalized (Our) & Personalized (Our)  \\
        \midrule
        \multirow{4}{*}{Beauty} & HR@1 & 0.0609 & 0.0604 & 0.0604 & \bm{$0.0613$} \\
        & HR@5 & 0.1558 & 0.1551 & 0.1544 & \bm{$0.1571^*$} \\
        & NDCG@5 & 0.1098 & 0.1089 & 0.1083 & \bm{$0.1101$} \\
        & HR@10 & \bm{$0.2326$} & 0.2310 & 0.2276 & 0.2281 \\
        & NDCG@10 & \bm{$0.1345$} & 0.1332 & 0.1318 & 0.1328 \\
        \midrule
        \multirow{4}{*}{Sports} & HR@1 & 0.0601 & 0.0587 & \bm{$0.0676^*$} & 0.0665 \\
        & HR@5 & 0.1620 & 0.1603 & 0.1783 & \bm{$0.1793$} \\
        & NDCG@5 & 0.1120 & 0.1104 & 0.1243 & \bm{$0.1244$} \\
        & HR@10 & 0.2417 & 0.2460 & 0.2653 & \bm{$0.2677^*$} \\
        & NDCG@10 & 0.1375 & 0.1378 & 0.1522 & \bm{$0.1527$} \\
        \midrule
        \multirow{4}{*}{Toys} & HR@1 & 0.0364 & 0.0367 & 0.0371 & \bm{$0.0379^*$} \\
        & HR@5 & \bm{$0.1138$} & 0.1110 & 0.1113 & 0.1133 \\
        & NDCG@5 & 0.0754 & 0.0742 & 0.0744 & \bm{$0.0757$} \\
        & HR@10 & 0.1840 & 0.1847 & 0.1842 & \bm{$0.1851$} \\
        & NDCG@10 & 0.0978 & 0.0978 & 0.0976 & \bm{$0.0986^*$} \\
        \bottomrule
    \end{tabular}
    \caption{Performance on user-item matching task using P5-B as the backbone. Bold numbers represent the best performance of each row. * indicates its performance is significantly better at $\bm{p < 0.05}$ than all other prompts.}
    \label{Table:matching_B}
\end{table}

\subsection{Performance Comparison}
\label{sec:performance_comparison}

We assess the effectiveness of the generated prompts in comparison to manually designed prompts using the test dataset for the three recommendation tasks. Specifically, we present the results for sequential recommendation with P5-S and P5-B in Table  \ref{Table:sequential_S} and Table \ref{Table:sequential_B}, the results for explanation generation using P5-S and P5-B in Table \ref{Table:explanation_S} and Table \ref{Table:explanation_B}, and the results for user-item matching using P5-S and P5-B in Table \ref{Table:matching_S} and Table \ref{Table:matching_B}, respectively. 
Each column stands for the performance of a type of prompt. Specifically, \textit{Seen manual} means the manual prompt used for the P5 pre-training, while \textit{Unseen manual} is not used for pre-training. \textit{Non-personalized} and \textit{Personalized} represent automated non-personalized and personalized prompts generated by our PAP-REC framework, respectively. Each row of the tables is the metric evaluation value on the test dataset, and several consequent rows are tested under the same dataset. 
The following observations are drawn from the results. 

For the sequential recommendation task, the automated prompts outperform the manual prompts under every experimental setting. Personalized and non-personalized prompts each achieved about half of the optimal results. 
Also, we can see that after applying our PAP-REC framework, P5-S and P5-B have great performance gains compared to using manual prompts.

For the explanation generation task, the generated prompts by PAP-REC also achieve all the best performance in most cases except for the \textit{Beauty} dataset when using the P5-B model. 
Under this setting, the \textit{seen manual} prompt in pre-training is about 1\% better on ROUGE-1 and ROUGE-L. However, similar to the precision-recall trade-off, the automated prompts both provide more than 5\% improvement on BLEU-4 and are much better than the \textit{unseen manual} prompts on all metrics in this model-dataset combination.

For the user-item matching task, 
some manual prompts can achieve the best performance on certain metrics. For example, the P5-S model achieves the best performance at HR@1 in the \textit{Beauty} dataset with the \textit{unseen manual} prompt; the \textit{seen manual} prompt slightly outperforms automated prompts in the \textit{Toys} dataset at HR@5. When applying P5-B on the \textit{Beauty} dataset, the \textit{seen manual} prompt performs the best on HR@10 and NDCG@10. However, either personalized prompts or non-personalized prompts beat manual prompts on the other metrics of various model-dataset combinations.

We provide some joint analysis across the three recommendation tasks. As mentioned in Section \ref{sec:surrogate_metrics}, we have opted to include only a subset of the approximate evaluation metrics to speed up the prompt generation process. The results indicate that the automatically generated prompts still consistently outperform their manually designed counterparts in the excluded metrics like HR@10 and NDCG@10 for the sequential recommendation. Also, PAP-REC can achieve better performances when using P5-S than P5-B as the backbone in some scenarios, as evidenced by metrics related to the sequential recommendation task. This is primarily attributed to the superior performance of P5-S over P5-B when employing manual prompts in these instances, given that the backbone model remains unchanged throughout the entire process. Therefore, the ultimate performance of PAP-REC is heavily influenced by the choice of the backbone model. Besides, when we compare the performance between automated personalized and non-personalized prompts, we can find that on the \textit{Sports} dataset, the non-personalized prompt tends to perform better, unlike the other two datasets. It may be because \textit{Sports} is a relatively sparser dataset, according to Table \ref{Table:dataset}, which makes the user-specific token more difficult to learn. 

\begin{table}[t]
    \centering
    \begin{tabular}{c|cccc}
        \toprule
         Prompt Position & HR@5 & NDCG@5 & HR@10 & NDCG@10 \\
         \midrule
         Prefix Only & 0.0204 & 0.0110 & 0.0330 & 0.0151 \\
         Prefix \& Suffix & 0.0503 & 0.0365 & 0.0652 & 0.0414\\
         Suffix Only & \bm{$0.0546$} & \bm{$0.0408$} & \bm{$0.0682$} & \bm{$0.0452$}\\
         \bottomrule
    \end{tabular}
    \caption{Sequential recommendation performance with different prompt positions on \textit{Beauty} dataset using P5-S as the backbone.}
    \label{Table:prompt_position}
\end{table}

\begin{figure}[t]
  \centering
  \includegraphics[width=0.7\linewidth]{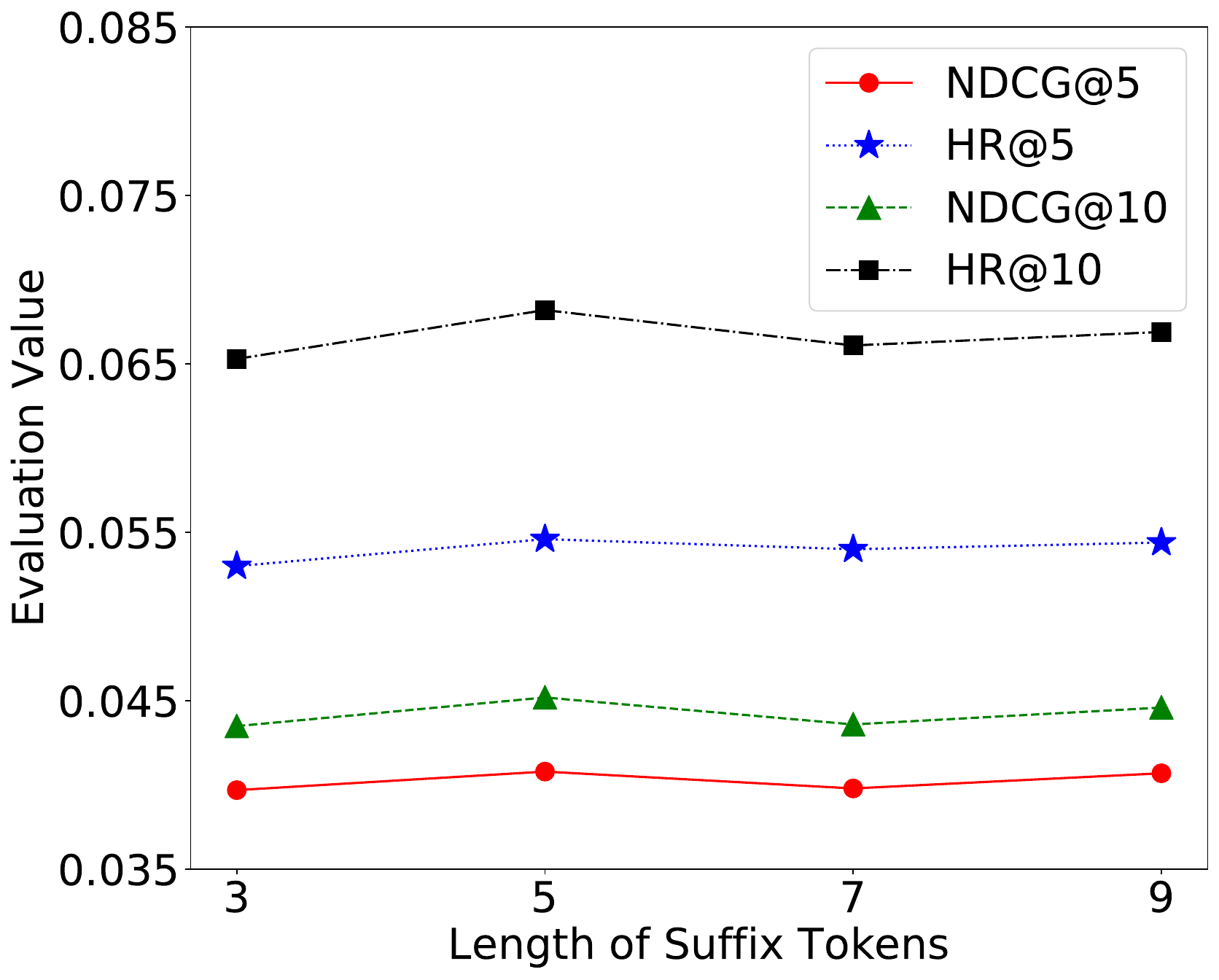}
  \caption{Sequential recommendation performance under different prompt lengths.}
  \label{fig:prompt_length}
\end{figure}

\begin{figure}
  \centering
  \includegraphics[width=0.7\linewidth]{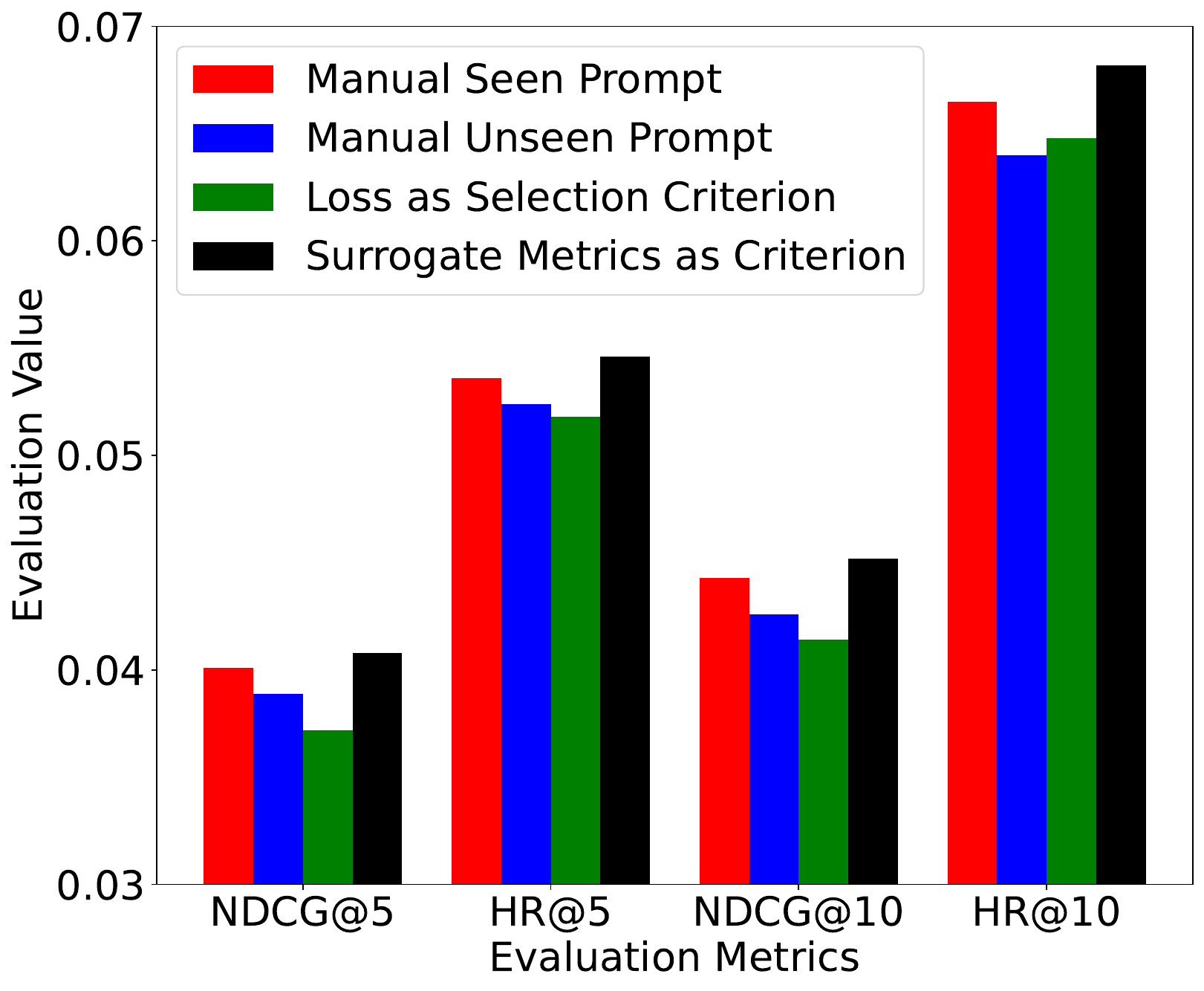}
  \caption{Ablation study on surrogate metrics $\widetilde{M}$ compared with loss as the selection criterion.}
  \label{fig:surrogate_metrics}
\end{figure}

\subsection{Analyses of the Automated Prompt Structure}
\label{sec:prompt_shape}

In this section, we examine the impact of various automated prompt structures on performance and analyze the reasons behind the impact.

\subsubsection{\bf Impact of trigger token positions} We first place the trigger tokens $[T_r]$ at different places in the prompt to see the impact. We name it as "prefix" for trigger tokens before \{\textit{task arg}\} and "suffix" for those after \{\textit{task arg}\}, both with the length of 5 as default. We conduct the experiment on the P5-S model and the \textit{Beauty} dataset, and the result is shown in Table \ref{Table:prompt_position}. We can see that the suffix prompt performs the best. A reason is that the manual prompt for P5 training describes the task detail after the \{\textit{task arg}\}, so the suffix prompt is necessary. Otherwise, the RLM may not understand the query as a sequential recommendation task.

\subsubsection{\bf Impact of the position of user-specific token} Considering that personalized automatically generated prompts contains a user-specific token $[T_u]$, we also need to determine the position of $[T_u]$. We test three situations, placing $[T_u]$ before \{\textit{task arg}\}, after $[T_r]$, and in-between them, on the P5-S and \textit{Beauty} model-dataset combination. Table \ref{Table:personalized_token} displays the performance. We can see that the RLM performs the best when $[T_u]$ is at the end of the prompt, which is consistent with Eq.\eqref{Eq:auto prompt}. Thus, we fix the $[T_u]$ position at the end as default.

\subsubsection{\bf Impact of the number of trigger tokens} Figure \ref{fig:prompt_length} shows the sequential recommendation performance of our automatically generated non-personalized prompts under different suffix prompt lengths $l = |\{[T_r]\}|$ when using \textit{Beauty} as dataset and P5-S as the backbone. We can see that when the length is too short, such as $l=3$ in the figure, the performance remains suboptimal compared to the best one, where $l=5$. The reason is that if the prompt is too short then it may not adequately describe the task and has little room to boost performance. Meanwhile, the performance decreases when we use more than 5 trigger tokens as the prompt. Since longer prompts lead to larger search space, it could take longer time and make it more difficult to reach the optimum, which is manifested as a degradation in the performance. Thus, we set $|[T_r]|=5$ as the default choice.

\begin{table}[t]
    \centering
    \begin{tabular}{c|cccc}
        \toprule
         Position of $\bm{[T_u]}$ & HR@5 & NDCG@5 & HR@10 & NDCG@10 \\
         \midrule
         $\bm{[T_u]}$\{\textit{task arg}\}\{$[T_r]$\} & 0.0482 & 0.0326 & 0.0639 & 0.0377\\
         \{\textit{task arg}\}$\bm{[T_u]}$\{$[T_r]$\} & 0.0474 & 0.0321 & 0.0635 & 0.0373\\
         \{\textit{task arg}\}\{$[T_r]$\}$\bm{[T_u]}$ & \bm{$0.0559$} & \bm{$0.0414$} & \bm{$0.0688$} & \bm{$0.0456$}\\
         \bottomrule
    \end{tabular}
    \caption{Sequential recommendation performance with different user-specific token positions in a personalized prompt when using P5-S as the backbone and \textit{Beauty} as the dataset.}
    \label{Table:personalized_token}
\end{table}

\subsection{Ablation Study on Surrogate Metrics}
\label{sec:ablation_study}

We have introduced the surrogate metrics in Section \ref{sec:surrogate_metrics} when selecting the best token from the top-k candidates. Surrogate metrics have brought significant efficiency improvement compared with using metrics as shown in Table \ref{Table:time comparison}. As for effectiveness, we leverage P5-S as the backbone and \textit{Beauty} as the dataset to illustrate the performance comparison between using loss functions and surrogate metrics in Figure \ref{fig:surrogate_metrics}. We can see that when using the loss as the token selection criterion to generate prompts, the final performance is almost consistently worse than P5 with manual prompts on all metrics, only better than the \textit{Unseen manual} prompt on HR@10; meanwhile, the prompt generated with surrogate metrics outperforms both baselines on all the metrics. Thus, the existing prompt generation methods using loss as the criterion in \cite{wallace-etal-2019-universal, autoprompt:emnlp20} cannot be effectively applied to recommendation language models. Besides, since most metrics of recommendation tasks are non-differentiable, we cannot directly use gradient-based methods for continuous prompt learning on the language model for recommendation.

\section{Conclusions and Future Work}
\label{sec:conclusions}

This paper proposes PAP-REC, an automatic personalized prompt generation framework for recommendation language models (RLM). We design surrogate metrics to solve the problems derived from recommendation-specific metrics, develop the iterative and alternative token update schedule to handle the challenges brought by inflating personalized tokens, and leverage the gradient to generate effective automated prompts. The automated prompts generated by our framework outperform manually designed prompts in most cases. Our framework can be applied to other sequence-to-sequence RLMs since the backbone model is frozen as a black box.

We will further extend our framework on several aspects in the future. As mentioned in Section \ref{sec:prompt_learning}, our research is orthogonal to the item ID learning, and there exists potential for a synergistic combination of these methods to complement each other. Also, in this work we designed a gradient-based prompt learning method for prompt generation. However, there exist other possible techniques for automated prompt generation, such as reinforcement learning and large language model-based methods, which are worth trying in the future. 
Finally, PAP-REC generates prompts for inference, but a more challenging line of potential research is to search for the best prompt for pre-training or fine-tuning, which may lead to new application scenarios for automated prompt generation and improve the performance of large language models. 

\newpage
% \balance
\bibliographystyle{unsrt}
\bibliography{paper.bib}

\end{document}